\documentstyle[12pt]{article}
\newcommand{\be}{\begin{equation}}
\newcommand{\ee}{\end{equation}}
\newcommand{\al}{\mbox{$\alpha$}}

\newcommand{\bi}[1]{\bibitem{#1}}
\newcommand{\fr}[2]{\frac{#1}{#2}}

\newcommand{\GD}{\mbox{$\tilde{G}$}}
\newcommand{\gf}{\mbox{$\gamma_{5}$}}

\newcommand{\Ima}{\mbox{Im}}

\begin{document}
\pagestyle{empty}
\normalsize
\begin{flushright}{UQAM-PHE/96-08\\September 1996}
\end{flushright}
\vspace{0.5cm}
\begin{center}{\Large \bf Radiative corrections to theta term in the left-right 
supersymmetric models}\\

\vspace{1.5cm}

{\bf M.E.Pospelov }\footnote{E-mail:pospelov@mercure.phy.uqam.ca} \\

\vspace{0.5cm}

{\em 
D\'epartement de Physique, Universit\'e du Qu\'ebec \`a
Montr\'eal,\\ Case Postale 8888, Succ. Centre-Ville, Montr\'eal,
Qu\'ebec, Canada, H3C 3P8}\\
and\\
{\em Budker Institute of Nuclear Physics, 630090 Novosibirsk, Russia}

\end{center}
\vspace{1.5cm}
\begin{abstract}
We calculate the radiative correction to the theta term in the generic 
left-right supersymmetric model due to the Kobayashi-Maskawa source of 
CP-violation. We found that the value of $\bar{\theta}$ is very sensitive to the 
relations between vacuum expectation values of bidoublet scalars 
$\langle\Phi_1\rangle=diag(\kappa_1,\,\kappa_1')$ and 
$\langle\Phi_2\rangle=diag(\kappa_2',\,\kappa_2)$. The minimal value of 
$\bar{\theta}$ in the model is found to be of order $10^{-9}$ for 
$\kappa_1/\kappa_2\sim 1$, $\kappa_1'=\kappa_2'=0$ in agreement with the 
experimental constraint without an axion mechanism or fine tuning. In other 
regions of the parameter space, the radiatively induced $\bar{\theta}$ gives 
unacceptably large contributions to the electric dipole moment of the neutron. 
\end{abstract}
\newpage

\pagestyle{plain}
\pagenumbering{arabic}

\section{Introduction} 

The strong CP problem is one of the most intriguing issues of modern particle 
physics. The additional term in the QCD Lagrangian 
\be
{\cal L}= \theta\fr{g^2_3}{16\pi^2} G^a_{\mu\nu}\GD^a_{\mu\nu} 
\ee
violates P and CP symmetries \cite{theta}. In the electroweak theory, the 
diagonalization of the quark mass matrices $M_u$ and $M_d$ involves chiral 
rotations and brings the additional contribution to the theta term:
\be
\bar{\theta}=\theta+arg(det M_uM_d)
\ee

The current experimental limits on the electric dipole moment (EDM) of the 
neutron put a severe constraint on the $\bar{\theta}$ parameter. The chiral 
algebra calculation of the neutron EDM induced by the theta term  \cite{CDVW} 
gives the following prediction:
\be
d_n\simeq 3.6\times10^{-16}\bar{\theta}\,e\cdot cm.
\ee
Together with the current neutron EDM constraints it implies the limit 
$\bar{\theta}<10^{-10}$. Bearing in mind other alternative ways to calculate EDM 
and the big diversity of the results (See, for example the review \cite{Chang}) 
we shall assume here the following milder limit for $\bar{\theta}$:
\be
\bar{\theta}<10^{-9}.
\label{eq:limit}
\ee

The extreme smallness of  $\bar{\theta}$ could be explained theoretically in 
different manners. The most popular solution for strong CP problem is to allow 
the dynamical relaxation of $\bar{\theta}$ through the axion mechanism 
\cite{PQ}. Since no axion, visible or invisible, is found so far, one has to 
consider other alternative ways to obtain naturally small $\bar{\theta}$ 
\cite{LR,NB}. 

\section{Radiative corrections to $\bar{\theta}$ }

In recent works Kuchimanchi \cite{K} and Mohapatra and Rasin \cite{MR1}, 
\cite{MR2} proposed a solution for the strong CP-problem in the framework of the 
supersymmetric models conserving parity. The theta parameter in the Lagrangian 
is simply set to zero above some scale $M_{W_R}$ where parity and CP are the 
exact symmetries of the theory. After spontaneous symmetry breaking, at the 
scale where $W_R$ becomes massive, the $\bar{\theta}$ parameter picks up no 
contribution from $arg(det M_uM_d)$. This is because the minimum of the 
superpotential corresponds to the real vacuum expectations values of scalar 
fields which leads to hermitean mass matrices \cite{K,MR1,MR2}. It does not 
mean, however, that the strong CP problem is solved; the theta term can be 
generated through radiative corrections if there is a CP-violating source in the 
theory. 

It is clear that to ensure these radiative corrections $\bar{\theta}_{rad}$ to 
satisfy the limit (\ref{eq:limit}) and thus to solve the CP-problem completely, 
one has to eliminate all extra sources of CP-violation beyond the 
Kobayashi-Maskawa (KM) phase. The latter provides a {\em minimal} content of 
CP-violation. If the contribution to $\bar{\theta}$ from KM phase happens to be 
large, this means that one cannot obtain the viable solution to the strong CP 
problem without fine tuning. This question was studied in the framework of pure 
SM \cite{EG,Kh}, where radiative corrections to $\bar{\theta}$ arise first in 
the order $\al_sG_F^2m_c^2m_s^2$ times the CP-odd KM invariant \cite{Kh}, and in 
the MSSM with the KM mechanism of CP-violation \cite{DGH} where the result also 
is found to be much smaller than $10^{-9}$. 

Here we address the same question to the generic left-right supersymmetric model 
and calculate the theta term unduced by the radiative corrections through the KM 
type of CP-violation. The simple estimate of the upper limit for $\bar{\theta}$, 
$\bar{\theta}<\fr{\al_s}{64\pi^3}
\Ima(V^*_{td}V_{tb}V^*_{cb}V_{cd})\times \,\log(M_{W_R}/M_{W_L})\sim 10^{-8}$, 
presented in the work \cite{MR2} is not satisfactory because it can predict the 
electric dipole moment of the neutron one order of magnitude above the present 
experimental limit. This estimate does not take into account the dependence of 
the quark masses which should be associated with the KM-type of CP-violation.
  
As usual, the potentially large CP-violating effects emerge through the one loop 
induced by quark-squark-gluino interaction. Following the works \cite{K,MR1,MR2} 
we take all new CP-violating phases specific for supersymmetric models to be 
equal to zero as the result of the parity conservation at $\Lambda_{GUT}$ scale:
\be
A=A^*;\; B=B^*;\; m_{\lambda_i}=m_{\lambda_i}^*;\; \mu_{ij}= \mu_{ij}^*.
\ee

The left-right symmetry imposed on the interaction of the quarks with Higgs 
bidoublets $\Phi_1$ and $\Phi_2$ requires the hermiticity of the Yukawa 
matrices:
\begin{eqnarray}
{\cal L}_Y=Y_1\bar{Q}_L \Phi_1 Q_R + Y_2 \bar{Q}_L \Phi_2 
Q_R\,+\,H.c.\nonumber\\
Y_1=Y_1^\dagger;\;\;Y_2=Y_2^\dagger
\end{eqnarray}
We do not specify here the particular content of the Higgs sector giving masses 
to $M_{W_R}$ in order to obtain the maximum of generality. The reality of the 
vacuum expecation values (VEV's) for Higgs bidoublets $\Phi_1$ and $\Phi_2$,
\be
\langle\Phi_1\rangle = 
\left(\begin{array}{cc}\kappa_1&0\\0&\kappa'_1\end{array}\right);\;
\langle\Phi_2\rangle = 
\left(\begin{array}{cc}\kappa'_2&0\\0&\kappa_2\end{array}\right),
\ee
corresponds to the minimum of the superpotential \cite{K,MR1,MR2}. It ensures 
the hermiticity of the mass matrices $M_u$ and $M_d$ and provides the same KM 
matrices for left- and right-handed charged currents. To get the simplest 
relations between mass matrices and Yukawa couplings and to avoid the problems 
with flavour changing neutral currents, we assume for the moment that 
$\kappa_1'=\kappa_2'=0$. Then $M_u$ and $M_d$ read as follows:
\be
M_u=\kappa_1Y_1\equiv \kappa_u\lambda_u;\;\;M_d=
\kappa_2Y_2^\dagger\equiv\kappa_d\lambda_d,
\ee
where $\kappa_u$ and $\kappa_d$, $\lambda_u$ and $\lambda_d$ are introduced from 
matter of convenience. As in the MSSM there is one additional free parameter, 
$\tan\beta=\kappa_u/\kappa_d$.

Let us now turn to the squark mass sector. The mass matrix for the down type 
squarks has the following general form:
\be
(\tilde{D}_L^\dagger\;
\tilde{D}_R^\dagger)
\left(
\begin{array}{cc}
m_L^2+c_u \lambda_u^2+ c_d \lambda_d^2&{\cal A}_d\\
{\cal A}_d^\dagger &m_R^2+c_u' \lambda_u^2+ c_d' \lambda_d^2 ,
\end{array}
\right)
\left(\begin{array}{c}
\tilde{D}_L\\
\tilde{D}_R
\end{array}\right),
\label{eq:mass}
\ee
where ${\cal A}_d=(A-\mu\tan\beta)(M_d+a_d\lambda_d^2M_d+
a_u\lambda_u^2M_d+a'_uM_d\lambda_u^2)$.\newline
The coefficients $c_u,\, c_u',\,c_d,\, c_d',\, a_d,\, a_u,\,a'_u$ appear either 
at the tree level or in the one-loop renormalization from $\Lambda_{GUT}$. The 
obvious requirement of the L-R symmetry is:
\be
m_L=m_R,\; c_d= c_d',\; c_u= c_u'\; a_u= a_u'. 
\label{eq:LR}
\ee
As a result the mass matrix (\ref{eq:mass}) differs from that of the MSSM where 
$c_u'=0$ and $a_u'=0$. The values of all these coefficients depend on many 
additional parameters and we simply assume here the following estimate: $ 
c_u\sim c_u'\sim m_{susy}^2(16\pi^2)^{-1}\ln(\Lambda_{GUT}^2/M_{W_R}^2)\sim 
{\cal O}(m_{susy}^2)$.

Let us now estimate the CP-violating mass term for quarks induced by the 
squark-gluino loop. The characteristic loop momenta are of order $m_{susy}$ and 
as a first approximation we can expand the propagators of squarks in series of 
the fermion U-quark Yukawa couplings. This expansion has the following simple 
form:
\be
(A-\mu \tan\beta)\sum_{n,m}\fr{c_u^nc_u'^m\left(V^
\dagger \lambda_u^{2n}
(VM_dV^\dagger+a_u\lambda_u^2VM_dV^\dagger
+a'_uVM_dV^\dagger \lambda_u^2)
\lambda_u^{2m}V\right)_{ii}}{(p^2-m_L^2)^{n+1}(p^2-m_R^2)^{m+1}},
\label{eq:nm}
\ee
where $V$ is the usual KM matrix and the subscript $ii$ denotes the projection 
on the initial flavour $i$. We have droped also all $c_d$-proportional terms as 
they are further suppressed by the D-quark Yukawa couplings. It is clear that if 
the conditions of the left-right symmetry (\ref{eq:LR}) are held, the expression 
(\ref{eq:nm}) is purely CP-conserving. In other words, in the mass eigenstate 
basis, the mixing matrices in the quark-squark-gluino couplings are identical 
for left- and right-handed particles and the CP-violating phase drops out at the 
one-loop level. However the further running of the mass parameters from the 
scale of parity violation down to the electroweak scale necessarily implies the 
departure from the exact relations (\ref{eq:LR}). As a result of that, the 
CP-violation can be developed, and the lowest-order term where it arises is 
$\lambda_t^4\lambda_c^2$. The explicit extraction of the CP-violating part from 
Eq.  (\ref{eq:nm}) for the external $d$-flavour leads to the following 
expression:
\begin{eqnarray}
(A-\mu 
\tan\beta)\Ima(V^*_{td}V_{tb}V^*_{cb}V_{cd})\lambda_c^2\lambda_t^4(m_b-m_s) 
\times\nonumber\\ \left[\fr{2a_uc_u(c_u-c_u')+2c_u^2(a_u-a_u')}{(p^2-m^2)^4}+
\fr{2a_uc_u^2(m_R^2-m^2_L)+c_u^2(c_u-c_u')}
{(p^2-m^2)^5} +\fr{c_u^3(m_R^2-m^2_L)}
{(p^2-m^2)^6}\right]
\end{eqnarray}
The differences between the coefficients $c_u$ and $c_u'$, $m_L$ and $m_R$ 
cannot be calculated without the knowledge of all masses below $M_{W_R}$. For 
our purposes, however, it is sufficient to use the reliable estimate for mass 
difference $m_L^2-m_R^2\sim
m_{susy}^2 6g_2^2(16\pi^2)^{-1}\ln(M_{W_R}^2/M_{W_L}^2)$ and similar relations 
for other coefficients. Combining together all these factors and performing the 
trivial integration, we arrive to the following form of the CP-violating quark 
masses:
\begin{eqnarray}
{\cal L}_5 \sim\Ima(V^*_{td}V_{tb}V^*_{cb}V_{cd})\fr{\al_s}{4\pi}
\fr{3\al_w}{2\pi}\mbox{ln}\fr{M^2_{W_R}}{M^2_{W_L}}\lambda_c^2
\lambda_t^4 \fr{m_{\tilde{G}}(A-\mu\tan\beta)}{m_{susy}^2}
F(m_{\tilde{G}^2}/m^2_{susy})\times\nonumber\\
\left[(m_b-m_s)\bar{d}i\gamma_5d +(m_d-m_b)\bar{s}i\gamma_5s + 
(m_s-m_d)\bar{b}i\gamma_5b \right]
\label{eq:g5} 
\end{eqnarray}
The exact form of the function $F$ is not important to us and we can take it 
$F\sim {\cal O}(1)$. All three CP-odd masses are suppressed by the square of the 
charm quark Yukawa coupling as it should be. To sufficient accuracy we can take 
also $\lambda_t^\simeq 1$ because no $\lambda_t$-expansion can be made. 

The analogous calculation of the radiatively induced CP-violating gluino mass 
term yields the following result:
\be
m_{\tilde{G}}-m_{\tilde{G}}*\sim 
i\Ima(V^*_{td}V_{tb}V^*_{cb}V_{cd})\fr{\al_s}{4\pi}
\fr{3\al_w}{2\pi}\mbox{ln}\fr{M^2_{W_R}}{M^2_{W_L}}
\fr{(A-\mu\tan\beta)m_b^2\lambda_c^2\lambda_s}{m_{susy}^2}
\ee
Due to the additional suppressions by the D-quark masses, this imaginary gluino 
mass gives just a negligible contribution to the theta-term. The main 
contribution to $\bar{\theta}$ comes from $\bar{d}i\gf d$-operator:
\be
\bar{\theta}\sim \Ima(V^*_{td}V_{tb}V^*_{cb}V_{cd})\fr{\al_s}{4\pi}
\fr{3\al_w}{2\pi}\mbox{ln}\fr{M^2_{W_R}}{M^2_{W_L}}
\fr{m_{\tilde{G}}(A-\mu\tan\beta)}{m_{susy}^2}
\lambda_c^2\fr{m_b}{m_d}
\label{eq:anal}
\ee

The interesting feature of this formula is a sort of "chiral" enhancement 
$m_b/m_d$ which is natural in the framework of the left-right model and simply 
impossible in MSSM where the chirality flip is always proportional to the mass 
of the external fermion. (This formula is valid only for the situation when the 
coefficient in front of $\bar{d}i\gamma_{5}d$ is much smaller than $m_d$). 
Substituting the numbers into (\ref{eq:anal}), we get the following estimate for 
the theta term developed in the generic left-right supersymmetric model with the 
KM-source of CP-violation:
\be
|\theta|\sim 10^{-9}\left\{\begin{array}{c}
                                     \tan\beta\; \;\;\;\mbox{for}\; 
\tan\beta\gg1\\
                                     {\cal O}(1)\;\;\;\; \mbox{for}\; 
\tan\beta\sim 1\\
                                     \tan^{-2}\beta\; \;\;\;\mbox{for} 
\;\tan\beta\ll 1
                         \end{array}\right.
\label{eq:answ}
\ee
When obtaining (\ref{eq:answ}) out of Eq. (\ref{eq:anal}), we took 
$\Ima(V^*_{td}V_{tb}V^*_{cb}V_{cd})\simeq 2.5\cdot 10^{-5}$; $m_{\tilde{G}}\sim 
|A|\sim |\mu| \sim m_{susy}$ and $\mbox{ln}(M^2_{W_R}/M^2_{W_L})\simeq 7$. 

\section{Discussion}

The common wisdom that the KM mechanism always gives the negligibly small 
contribution to the CP-violating flavour-conserving observables apparently is 
not true in the case of the left-right supersymmetric model. We have shown that 
the radiative corrections to the $\bar{\theta}$-parameter in the generic 
left-right supersymmetric model are large, just about the edge of the current 
experimental constraint. The only contribution to theta term comes from the 
radiative corrections to the $d$-quark mass. The main difference of our answer 
(\ref{eq:answ}) in comparison with the simple estimate quoted in \cite{MR2} is 
in the additional multiplier $\lambda_c^2m_b/m_d$ which is of the order 
$5\cdot10^{-2}$ for $\kappa_1\sim \kappa_2$ and $ \kappa_1'= \kappa_2'=0$. In 
this domain of the parameter space, the radiatively induced $\bar{\theta}$ is 
hard but not impossible to reconcile with the current experimental limit. One 
way for that would be to make the ratio 
$m_{\tilde{G}}(A-\mu\tan\beta)/m_{susy}^2$ reasonably small, of order $10^{-1}$.  

It turns out that the value of  $\bar{\theta}$ is very sensitive to the 
relations between different VEV's of the model. Thus, the Eq. (\ref{eq:answ}) 
suggests that both small and large $\kappa_1/\kappa_2$ are almost excluded. In 
the more general formulation of the model $\kappa_1',$ $ \kappa_2'$ also differ 
from zero. In that case we observe another contribution to $\bar{\theta}$ which 
is suppressed only by the first power of the charm quark Yukawa coupling. This 
contribution comes from the cubic term $a_u(\lambda_u^2M_d+M_d\lambda_u^2)\simeq 
a_u(\kappa'_1/\kappa^3)M_uM_dM_u$ in the mixing of the left- and right-handed 
squarks. The overall factor $\lambda_c^2$ in the estimate (\ref{eq:anal}) is 
then substituted for $\lambda_c\kappa'_1/\kappa_1$. To keep this contribution in 
 agreement with the experimental limit, one has to assume that 
$\kappa'_1/\kappa_1<10^{-2}$. This constraint is held even in the limit  of very 
large $m_{susy}$ and $M_{W_{R}}$ where many other phenomenological constraints 
(such as the flavour changing neutral currents) are trivially satisfied. In the 
limit $M_{W_{R}}\longrightarrow\infty$ the squark mass matrix keeps the 
nonvanishing remnants of the left-right symmetry resembling the case of the 
supersymmetric $SO(10)$ models \cite{SO10} where the radiative corrections to 
$\theta$ are also known to be large (the last Ref. in \cite{SO10}).  

If the strong CP-problem is cured by the axion, CP-violating mass term 
(\ref{eq:g5}) has no effect on the physical observables. The EDM of the neutron, 
in this case, originates from operators of dimension bigger than 4, such as EDM 
of quarks, color EDM of quarks, etc. We give a crude estimate for the EDM of the 
neutron using the size of coefficient in front of $\bar{d}\gamma_5d$ in 
(\ref{eq:g5}) multiplied by $e/m_{susy}^2$ which is of the order $10^{-29}e\cdot 
cm$ for $m_{susy}$ taken close to electroweak scale. 

I would like to thank C. Burgess, G. Couture, M. Frank, C. Hamzaoui, H. K\"onig 
and A. Zhitnitsky for many helpful discussions. This work is supported by NATO 
Science Fellowship, N.S.E.R.C., grant \#  189 630 and Russian Foundation for 
Basic Research, grant \# 95-02-04436-a.

\end{document}